\documentclass[conference]{IEEEtran}
\IEEEoverridecommandlockouts

\usepackage{cite}
\usepackage{amsmath,amssymb,amsfonts}
\usepackage{algorithm}
\usepackage{algpseudocode}
\usepackage{graphicx}
\usepackage{booktabs}
\usepackage{multirow}
\usepackage{textcomp}
\usepackage{subcaption}
\usepackage{hyperref}
\usepackage{xcolor}
\def\BibTeX{{\rm B\kern-.05em{\sc i\kern-.025em b}\kern-.08em
    T\kern-.1667em\lower.7ex\hbox{E}\kern-.125emX}}
\begin{document}

\title{SimQFL: A Quantum Federated Learning Simulator with Real-Time Visualization}

\author{
        \IEEEauthorblockN{
        Ratun Rahman\IEEEauthorrefmark{0}, Atit Pokharel\IEEEauthorrefmark{0}, Md Raihan Uddin\IEEEauthorrefmark{0}, and Dinh C. Nguyen\IEEEauthorrefmark{0}
	}

	\IEEEauthorblockA{
	\IEEEauthorrefmark{0}Department of Electrical and Computer Engineering, The University of Alabama in Huntsville, AL, USA 
	}
 
 Emails: rr0110@uah.edu, ap1284@uah.edu, mu0016@uah.edu, dinh.nguyen@uah.edu}
	\markboth{}%
	{}

\maketitle

\begin{abstract}
Quantum federated learning (QFL) is an emerging field that has the potential to revolutionize computation by taking advantage of quantum physics concepts in a distributed machine learning (ML) environment. However, the majority of available quantum simulators are primarily built for general quantum circuit simulation and do not include integrated support for machine learning tasks such as training, evaluation, and iterative optimization. Furthermore, designing and assessing quantum learning algorithms is still a difficult and resource-intensive task. Real-time updates are essential for observing model convergence, debugging quantum circuits, and making conscious choices during training with the use of limited resources. Furthermore, most current simulators fail to support the integration of user-specific data for training purposes, undermining the main purpose of using a simulator. In this study, we introduce SimQFL, a customized simulator that simplifies and accelerates QFL experiments in quantum network applications.  SimQFL supports real-time, epoch-wise output development and visualization, allowing researchers to monitor the process of learning across each training round. Furthermore, SimQFL offers an intuitive and visually appealing interface that facilitates ease of use and seamless execution. Users can customize key variables such as the number of epochs, learning rates, number of clients, and quantum hyperparameters such as qubits and quantum layers, making the simulator suitable for various QFL applications.  The system gives immediate feedback following each epoch by showing intermediate outcomes and dynamically illustrating learning curves. User can also upload their own data and check the simulation results using QFL without any fundamental knowledge. SimQFL is a practical and interactive platform enabling academics and developers to prototype, analyze, and tune quantum neural networks with greater transparency and control in distributed quantum networks.
\end{abstract}


\section{Introduction}
Simulators are important in both conventional and quantum computing research because they enable algorithm development, testing, and validation in controlled and repeatable environments.  In the field of federated learning (FL), simulators allow academics to experiment with collaborative training methodologies without the requirement for a large-scale distributed infrastructure.  Open-source FL platforms, such as OpenFL \cite{reina2021openfl} and PeerFL\cite{luqman2024peerfl}, have emerged as important tools, supporting a wide range of training scenarios and facilitating rigorous evaluation of federated algorithms across a variety of network, client, and data settings. Similarly, simulators such as Fujitsu's 40-qubit system \cite{imamura2022mpiqulacs} and platforms described at QSim 2024 \cite{qsim2024conference} are critical environments for developing quantum circuits and evaluating quantum algorithms on classical hardware.  These simulators not only decrease the barrier to testing but also provide insights into algorithm behavior, resource efficiency, and performance bottlenecks. 

Quantum machine learning (QML) investigates the integration of quantum computing principles with machine learning approaches in order to provide computational benefits such as faster processing, richer data representations, and improved model performance \cite{khurana2024quantum}.  Using quantum features such as superposition and entanglement, QML has brought new paradigms in data encoding, neural topologies, and optimization. However, one major problem with QML is that i) \textit{it requires data centralization}, which creates data privacy concerns \cite{pokharel2025quantum}, and ii) \textit{centralized quantum processing}, which requires significant quantum resources at the central server \cite{innan2024fedqnn}. 

By addressing these drawbacks with decentralized, privacy-preserving learning processes called federated learning (FL), quantum federated learning (QFL) shows significant potential as an extension to QML. Without disclosing raw data or needing centralized quantum processing, QFL allows many quantum clients to train local quantum models on confidential datasets and collaboratively update a shared global model \cite{ren2023towards, rahman2025sporadic}. This configuration offers built-in benefits in data security, scalability, and error tolerance while significantly reducing the requirement for centralized quantum hardware. By modeling diverse client behaviors, multiple data distributions (including non-IID settings), and varied quantum configurations (such as the number of qubits or circuit depth), a well-designed QFL simulator can facilitate experimentation with these decentralized quantum models \cite{qiao2024transitioning}. Such a simulator helps researchers evaluate performance trade-offs between privacy, accuracy, and utilization of resources, in addition to assisting with algorithmic development. 
However, current simulators exhibit two practical challenges. 

\textbf{i) Lacking integrated support for QFL:} QFL platforms lack quantum features like qubit encoding, variational quantum layers, or support for hybrid quantum-classical models, despite their great adaptability in conventional settings \cite{park2025entanglement}. In contrast, general-purpose quantum simulators emphasize circuit reliability and quantum gates but provide minimal support for multi-client synchronization, model training, or optimization \cite{gurung2025performance}. As a result, researchers studying QFL are frequently obliged to improvise tools or experiment with non-interactive scripts. This brings up the first important question: a) \textit{How can researchers use only classical hardware to efficiently build and examine the training dynamics of QML models in a federated environment?}

\textbf{ii) Providing a visually appealing interface and real-time insight:} In addition, current simulators are inadequate in terms of ease of execution and visually appealing during training. Subsequently, conventional FL or quantum simulation tools hardly enable real-time visualization, which is crucial for managing computing resources, detecting convergence, and troubleshooting quantum layers. For example, methods that investigate non-IID data impacts in QFL, such as the one by Zhao et al. \cite{zhao2023non}, depend on post-training assessments without ongoing observation. Comparably, privacy-preserving QFL frameworks that emphasize cryptographic integration, as MQFL by Dutta et al. \cite{dutta2024mqfl}, lack visual interpretability during training. This raises a second crucial question: b) \textit{How can researchers use visually appealing, real-time insights into performance and learning behavior to examine and refine QFL models interactively?}

\textbf{iii) Personal data training:} Furthermore, most existing QFL and simulation systems do not allow users to upload and use their own datasets for personalized model training. This shortcoming greatly affects the usefulness of such tools, especially when the reason for using a simulator is to apply a learning method without fundamental knowledge. In order to create reliable and broadly applicable QFL systems, simulators must be able to incorporate user-specific data. Otherwise, they are unable to represent client-specific distributions, preferences, or noise characteristics. As a result, this raises a third key question: c) \textit{In order to facilitate user-centric experimentation in QFL research, how can simulators facilitate the smooth and private integration of personal datasets?}

\textbf{Main Contributions:} To answer these significant questions, \textit{we introduce SimQFL, a simulator for quantum federated learning that has integrated support for real-time visualization and distributed quantum model training}. The primary goal is to provide a useful, adaptable, and interactive platform that lets developers and researchers test, prototype, and evaluate QFL algorithms without requiring complicated distributed systems or quantum tools. This paper's main contributions are listed below.
\begin{itemize}
    \item We present \textit{SimQFL}, a novel simulator that enables decentralized quantum model training across multiple simulated clients by integrating the concepts of quantum machine learning and federated learning for a more practical deployment. The simulator is visually appealing and easy to use for seamless execution. 
    \item SimQFL does not require a deep understanding of quantum federated learning; users may upload and train on their own datasets.  This feature greatly increases accessibility, allowing researchers with various amounts of experience to conduct custom QFL training.
    \item We provide a fully flexible framework that allows users to change parameters such as the number of clients, quantum layers, qubits, learning rate, data distributions (IID and non-IID), and encoding schemes, making SimQFL suitable for a wide range of QFL research applications.
\end{itemize}



\section{Related Works}

66;; We provide a summary of related works from various aspects relevant to our QFL development.

Federated Learning (FL) has attracted a lot of interest for allowing group model training free from centralized data collection \cite{rahman2025electrical, rahman2025multimodal, rahman2024multimodal}. Many open-source systems have been created to support FL research. Covering the top open-source FL frameworks, OpenFL \cite{reina2021openfl} explores their respective benefits and drawbacks. PeerFL \cite{luqman2024peerfl} also presents a simulator made for peer-to-peer FL at scale that integrates FL tools with the NS3 network simulator to replicate heterogeneous device experiments. These simulators provide the means to test FL algorithms and system behaviors in many contexts.

Essential tools for imitating quantum systems and enabling the evolution of quantum algorithms are quantum simulators. Using a CPU-based state vector approach to attain great computational efficiency, Fujitsu's 40-qubit quantum simulator \cite{imamura2022mpiqulacs} marks a major development in this field. Furthermore, providing a forum for debating the most recent advancements in quantum simulation, the QSim 2024 conference \cite{qsim2024conference} brings together professionals from several fields to investigate both analog and digital methods.

To handle challenging computational tasks \cite{pokharel2024electrocardiogram}, QML investigates the integration of quantum computing concepts with machine learning methods. Providing a thorough overview of QML, including basic ideas, techniques, and statistical learning theories relevant to the subject, Wang and Liu \cite{wang2024comprehensive}. Their work shows how, using quantum advantages, QML may transform data processing and analysis.
Several architectures and approaches have been suggested in the quickly developing subject of QFL to improve the applicability and efficiency of distributed quantum machine learning. By providing a customizable and real-time visualizing platform for QFL experiments, the proposed SimQFL simulator aligns with and expands existing developments.

\begin{figure}
    \centering
    \includegraphics[width=0.99\linewidth]{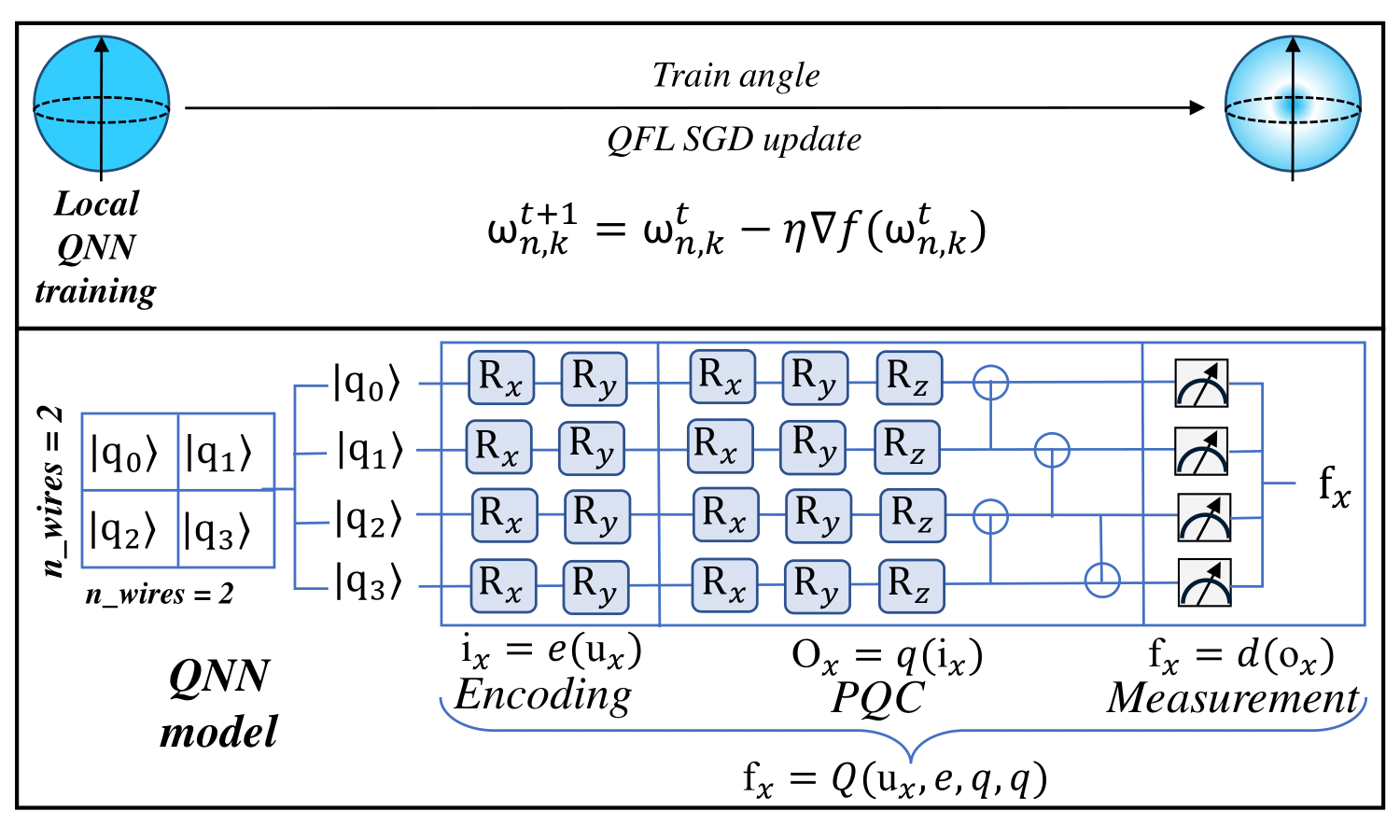}
    \caption{Overview of the QFL architecture with $N$ quantum clients and a classical server within the QFL framework. Each client encodes conventional input into quantum states and is trained using local Quantum Machine Learning (QML) models. The server aggregates the parameters of the local models and updates the global model parameters for the next round. }
    \label{fig: overview}
\end{figure}

Using genomic datasets on quantum simulators, Pokhrel et al. \cite{pokhrel2024quantum} presented a data-encoding-driven QFL framework, showing its feasibility. Their work offers a proof of concept showing good results and highlights the need for data encoding in QFL. SimQFL expands on this by letting users choose data encoding techniques, therefore enabling a wider spectrum of applications.

In QFL, Zhao et al. \cite{zhao2023non} investigated the difficulties of non-independent and identically distributed (non-IID) data. Using both theoretical and numerical studies, they suggested a structure to handle the non-IID problem. SimQFL uses procedures to replicate non-IID data distributions, therefore allowing researchers to investigate their effects on QFL models with relative efficiency.

Using quantum computing to improve performance while preserving data privacy via totally homomorphic encryption, Dutta et al. \cite{dutta2024mqfl} presented a multimodal QFL framework. SimQFL enhances this by giving a forum for prototyping and analysis of privacy-preserving QFL models with real-time feedback.

Chehimi et al. \cite{chehimi2023foundations} investigated the possibilities and difficulties of QFL over classical and quantum networks holistically. They pointed out important elements and special difficulties in implementing QFL, thereby providing fresh ideas and study paths. SimQFL provides an interactive and realistic environment for developing and tuning quantum neural networks in distributed contexts, therefore addressing some of these difficulties.

Emphasizing cooperative training without sharing local data, Yu et al. \cite{yu2022quantum} presented a QFL architecture for distributed quantum networks. They developed a quantum gradient descent approach for local model training and techniques for data information extraction into quantum states. SimQFL extends this by offering adjustable quantum hyperparameters, such as qubits and quantum layers, enabling flexible experimentation with related collaborative training methodologies.

Diadamo et al. \cite{diadamo2021qunetsim}, designed to ease research and instructional activities at the quantum network level, is a Python-based quantum network simulator. It is fit for the testing and development of network-layer quantum protocols since it offers real-time simulation features and supports several quantum networking architectures. QuNetSim mostly concentrates on quantum communication and lacks built-in federated learning capabilities, even if it offers interactive real-time simulations. By combining real-time visualization and numerous customization choices, especially for QFL experiments, our suggested SimQFL explicitly expands such simulation ideas into QFL environments.

Finally, SimQFL combines and expands current QFL techniques by providing a customizable, real-time visualization tool supporting several data distributions and quantum hyperparameter configurations, enabling thorough investigation and analysis in distributed quantum machine learning.


\begin{table*}[htbp]
\centering
\caption{Comparison of Existing Works and Proposed QFL Simulator (SimQFL)}
\renewcommand{\arraystretch}{1.3}
\begin{tabular}{|p{2.8cm}|c|c|c|c|c|p{4cm}|}
\hline
\textbf{Works/Features} & \textbf{FL} & \textbf{Quantum} & \textbf{Real-time Vis.} & \textbf{Customization} & \textbf{Interactive} & \textbf{Contribution} \\ \hline

\textbf{SimQFL (Proposed)} & \checkmark & \checkmark & \checkmark & High & \checkmark & Customizable, interactive QFL simulator with real-time visualization \\ \hline

QuNetSim \cite{diadamo2021qunetsim} & $\times$ & \checkmark & \checkmark & Moderate & Partial & Real-time quantum network simulation with extensive quantum protocols \\ \hline

Apheris AI \cite{reina2021openfl} & \checkmark & $\times$ & $\times$ & Moderate & $\times$ & Comprehensive classical FL framework comparison \\ \hline

PeerFL \cite{luqman2024peerfl} & \checkmark & $\times$ & Partial & High & Partial & Peer-to-peer federated learning simulation at scale \\ \hline

Fujitsu Simulator \cite{imamura2022mpiqulacs} & $\times$ & \checkmark & Partial & Moderate & Partial & High-performance quantum state simulation \\ \hline

Pokhrel et al. \cite{pokhrel2024quantum} & \checkmark & \checkmark & $\times$ & Moderate & $\times$ & Data-encoding-driven QFL using quantum simulators \\ \hline

Dutta et al. \cite{dutta2024mqfl} & \checkmark & \checkmark & $\times$ & Limited & $\times$ & Multimodal QFL with homomorphic encryption \\ \hline

\end{tabular}
\label{tab:comparison}
\end{table*}

\section{QFL Framework}
The QFL framework involves $N$ distributed clients collaboratively training a shared quantum neural network (QNN) model over $R$ global communication rounds. In each global round $r \in \{1, \dots, R\}$, clients $n \in \{1, \dots, N\}$ perform $K$ local training epochs on their private datasets $D_n$ without sharing raw data. After local training, each client sends its updated model parameters to a central server, which aggregates them to update the global model.

\noindent
\textbf{Quantum Encoding:}
Each client $n$ at global round $r$ maps classical input data $w \in \mathbb{R}^d$ into a quantum state using amplitude encoding. This method efficiently embeds classical vectors into quantum states by encoding their components into the amplitudes of computational basis states. For a $2^{Q_n}$-dimensional input vector, only $Q_n$ qubits are needed, providing the qubit usage efficiency.
Given a  input classical vector $w \in \mathbb{R}^{2^{Q_n}}$, its amplitude-encoded quantum state is
\begin{equation}
|\psi_{\text{enc}}(w)\rangle = \frac{1}{\|w\|} \sum_{i=0}^{2^{Q_n}-1} w_i |i\rangle,
\label{eq:encoding_amplitude}
\end{equation}
where $\|w\|$ is the $\ell_2$-norm of the vector, and $|i\rangle$ denotes the computational basis states.
This encoding can be represented via a unitary transformation applied to the initial state $|0\rangle^{\otimes Q_n}$, written as 
$|\psi_{\text{enc}}(w)\rangle = U_{\text{enc}}(w) |0\rangle^{\otimes Q_n}$.  
The unitary $U_{\text{enc}}(w)$ prepares the amplitudes according to the classical vector. The resulting state is passed to PQC for local training.

\noindent
\textbf{Local Model Training:}
Each client applies a Parameterized Quantum Circuit (PQC) $U(\omega_{n,r}^{k})$ to the encoded state $|\psi_{\text{enc}}(w)\rangle$. The PQC consists of $L$ layers, where each layer combines trainable single-qubit rotation gates (e.g., $R_x$, $R_y$, $R_z$) with fixed entangling gates (e.g., CNOT) to enable both local transformations and inter-qubit interactions. The state evolution can be expressed by a unitary operator as
\begin{equation}
U(\omega_{n,r}^{k}) = \prod_{d=1}^L U_d(\omega_{n,r,d}^{k}),
\label{eq:pqc_unitary}
\end{equation}
where $U_d(\omega_{n,r,d}^{k})$ is given by $\exp\left(-i \frac{\omega_{n,r,d}^{k}}{2} G_d\right)$ with $G_d$ being a Hermitian generator, typically a Pauli string.
The resulting output state is given by
\begin{equation}
|\psi_{\text{out}}(w, \omega_{n,r}^{k})\rangle = U(\omega_{n,r}^{k}) |\psi_{\text{enc}}(w)\rangle.
\label{eq:pqc_output_state}
\end{equation}
The measurement is performed using a Hermitian observable $O$ (e.g., Pauli-Z), producing a classical expectation value that can be expressed as
\begin{equation}
f(w, \omega_{n,r}^{k}) = \langle \psi_{\text{out}}(w, \omega_{n,r}^{k}) | O | \psi_{\text{out}}(w, \omega_{n,r}^{k}) \rangle.
\label{eq:measurement_expectation}
\end{equation}
Typically, this measurement is performed over $M$ shots so that $\hat{f}_{n,r}^{k}(w, \omega_{n,r}^{k})$ is given by $\frac{1}{M} \sum_{j=1}^{M} H_j$, where $H_j$ are measurement outcomes. The loss for a data point $(w, y)$ is computed as $\ell(y, \hat{f}_{n,r}^{k}(w, \omega_{n,r}^{k}))$.

For each epoch $k$, a mini-batch $B_{n}^{k} \subseteq D_n$ of size $|B_{n}^{k}|$ is sampled from the local dataset. The mini-batch loss is defined as
\begin{equation}
L_{n,r}^{k,B}(\omega_{n,r}^{k}) = \frac{1}{|B_{n}^{k}|} \sum_{(w, y) \in B_{n}^{k}} \ell(y, \hat{f}_{n,r}^{k}(w, \omega_{n,r}^{k})).
\label{eq:batch_loss}
\end{equation}

Gradients are computed using the parameter shift rule. For parameter $\omega_{n,r,d}^{k}$, the derivative can be expressed as
\begin{equation}
\frac{\partial f(w, \omega_{n,r}^{k})}{\partial \omega_{n,r,d}^{k}} = \frac{1}{2} \left[ f(w, \omega_{n,r}^{k} + \frac{\pi}{2} e_d) - f(w, \omega_{n,r}^{k} - \frac{\pi}{2} e_d) \right],
\label{eq:parameter_shift}
\end{equation}
where $e_d$ is a unit vector in the $d$-th direction. The full mini-batch gradient is denoted by $\nabla_{\omega} L_{n,r}^{k,B}(\omega_{n,r}^{k})$, and the parameters are updated as follows
\begin{equation}
\omega_{n,r}^{k+1} = \omega_{n,r}^{k} - \eta \nabla_{\omega} L_{n,r}^{k,B}(\omega_{n,r}^{k}),
\label{eq:parameter_update}
\end{equation}
where $\eta$ is the learning rate. After $K$ local training epochs, the final updated parameters in a client are denoted as $\omega_{n}^{(r)} = \omega_{n,r}^{K}$.

\noindent
\textbf{Federated Aggregation:}
At the start of each round $r$, the server broadcasts the global parameters $\boldsymbol{\omega}_{\text{global}}^{(r)}$ to all clients, who initialize $\boldsymbol{\omega}_{n,r}^{k=0} = \boldsymbol{\omega}_{\text{global}}^{(r)}$. After local training, clients send $\boldsymbol{\omega}_{n}^{(r)}$ to the server, and the aggregated model is computed as
\begin{equation}
\boldsymbol{\omega}_{\text{global}}^{(r+1)} = \frac{1}{D} \sum_{n=1}^{N} |D_n| \boldsymbol{\omega}_{n}^{(r)},
\label{eq:aggregation}
\end{equation}
where $D = \sum_{n=1}^{N} |D_n|$ weights contributions by dataset size. The updated global parameters are then redistributed for the next round, continuing until convergence or reaching the final global round $R$.

\section{Architecture}
The main objective of SimQFL is to offer an end-to-end simulation framework for interactive, fully configurable experiments with QFL using Flask.  The simulator is intended to replicate the actions of many quantum clients that use FL principles to train quantum machine learning models collaboratively.  Without requiring actual quantum devices or physical federated infrastructure, it allows researchers to examine QFL under various conditions, including alternative data distributions, quantum configurations, and optimization techniques.  With the help of SimQFL, users can set up training settings, view real-time model performance, and obtain a profound understanding of how decentralized quantum systems learn.  The simulator is also an effective instructional tool, offering step-by-step feedback throughout training cycles.

\begin{figure*}
    \centering
    \footnotesize

    \begin{subfigure}[t]{0.48\linewidth}
        \centering
        \includegraphics[width=\linewidth]{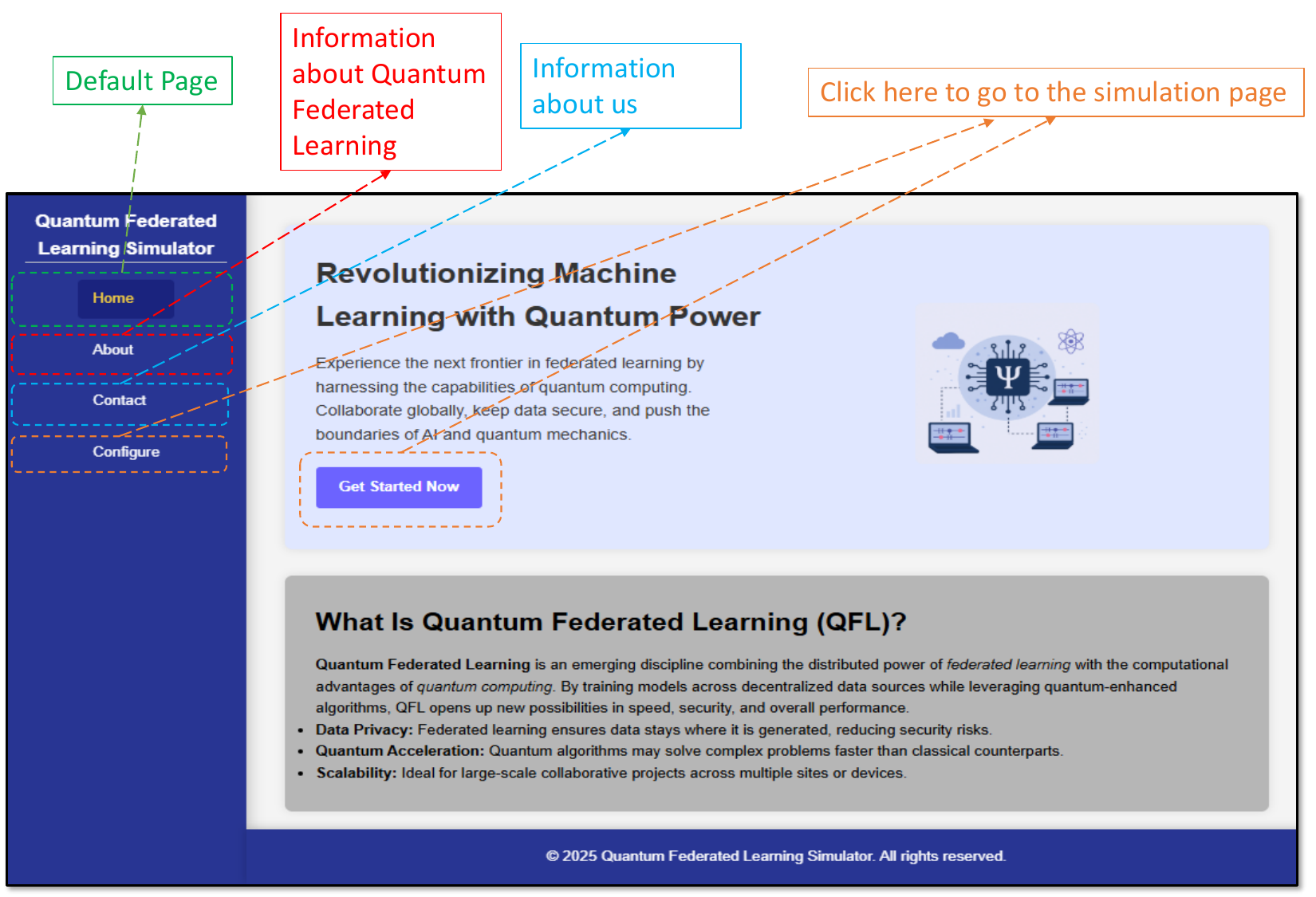}
        \caption{Homepage screen.}
        \label{fig:img1}
    \end{subfigure}
    \hfill
    \begin{subfigure}[t]{0.48\linewidth}
        \centering
        \includegraphics[width=\linewidth]{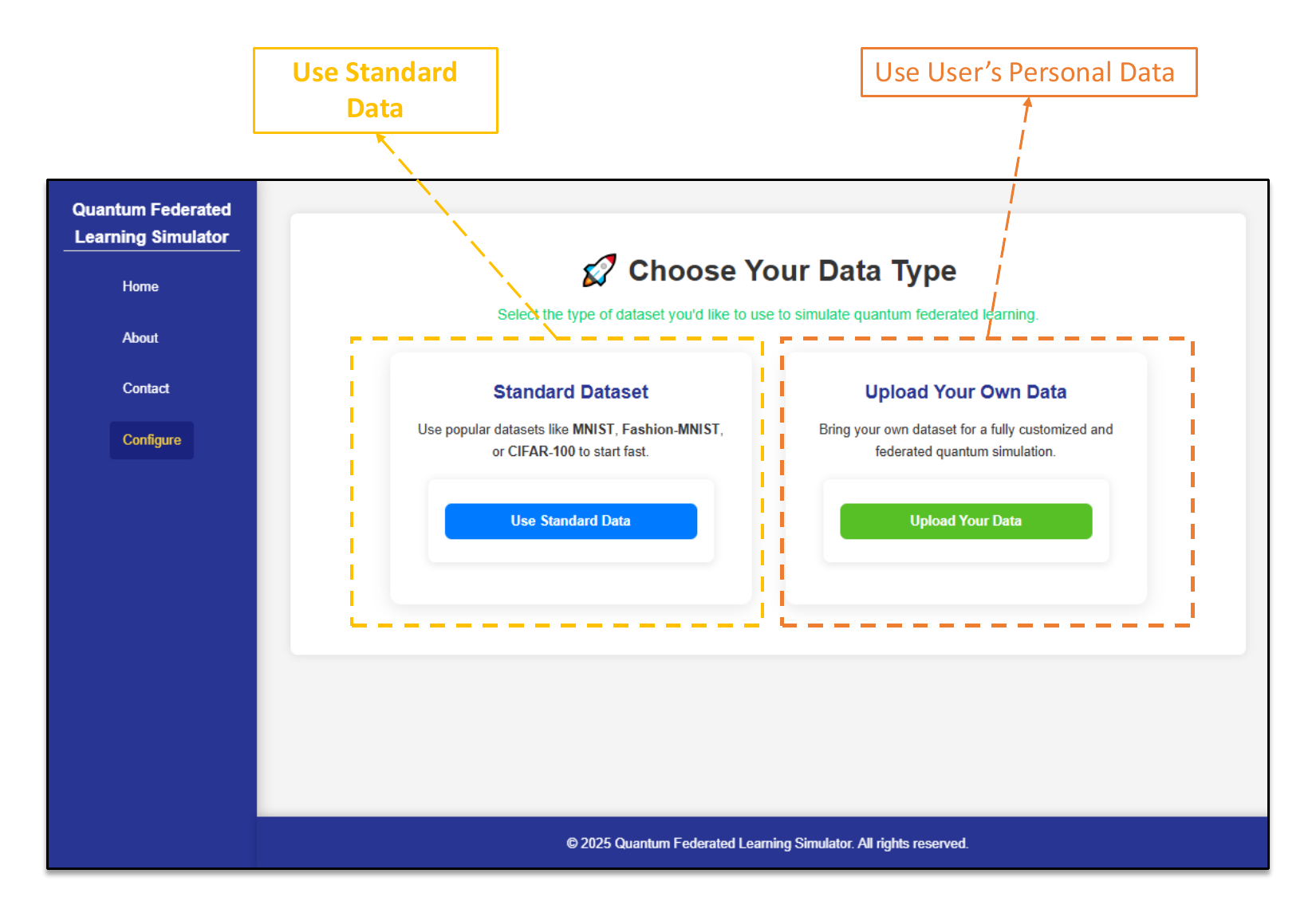}
        \caption{Configuration screen.}
        \label{fig:img1.5}
    \end{subfigure}

    \vspace{5pt}

    \begin{subfigure}[t]{0.48\linewidth}
        \centering
        \includegraphics[width=\linewidth]{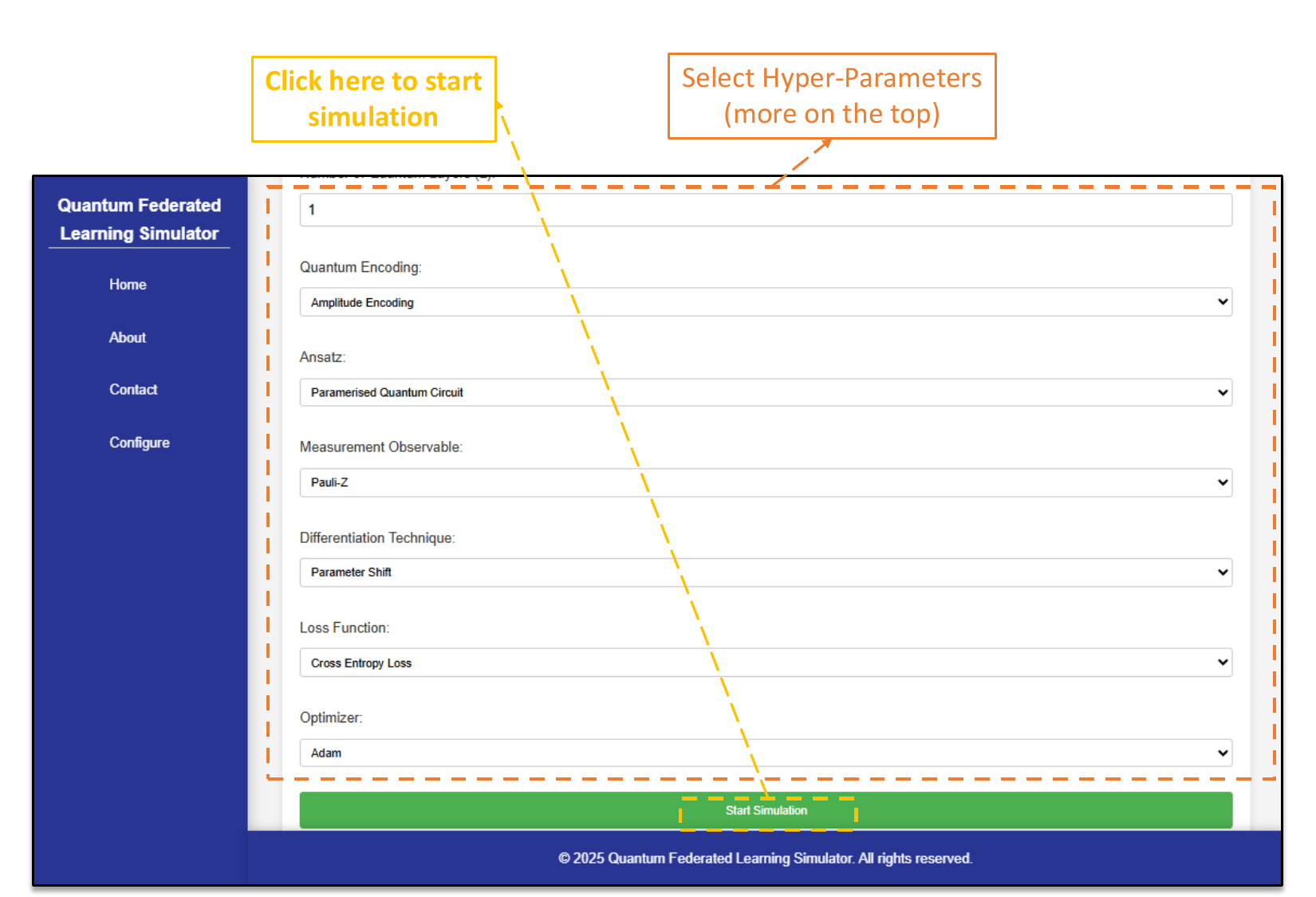}
        \caption{Simulation screen (Standard dataset).}
        \label{fig:img2}
    \end{subfigure}
    \hfill
    \begin{subfigure}[t]{0.48\linewidth}
        \centering
        \includegraphics[width=\linewidth]{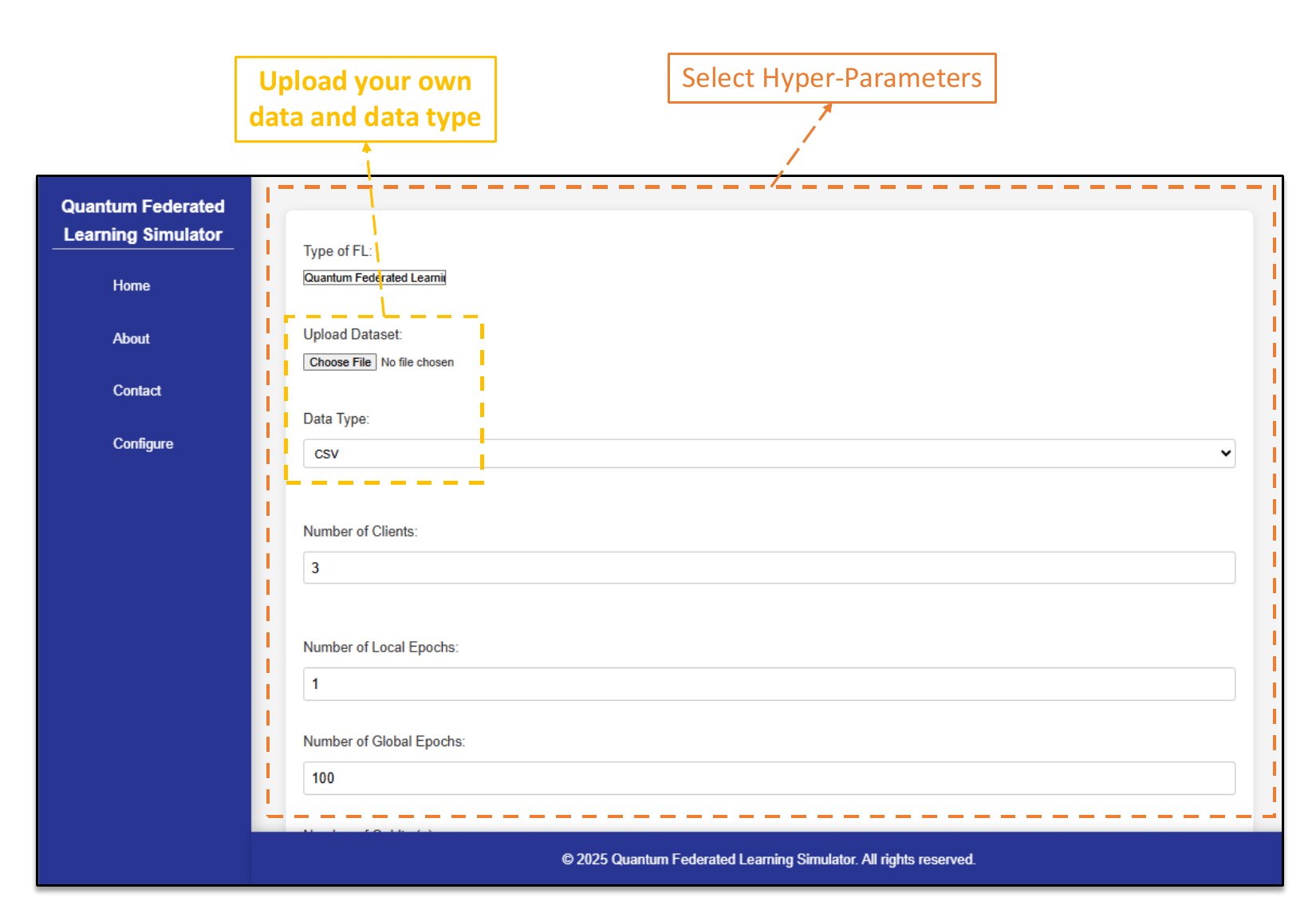}
        \caption{Simulation screen (Upload your own dataset).}
        \label{fig:img2.5}
    \end{subfigure}

    \vspace{5pt}

    \begin{subfigure}[t]{0.33\linewidth}
        \centering
        \includegraphics[width=\linewidth]{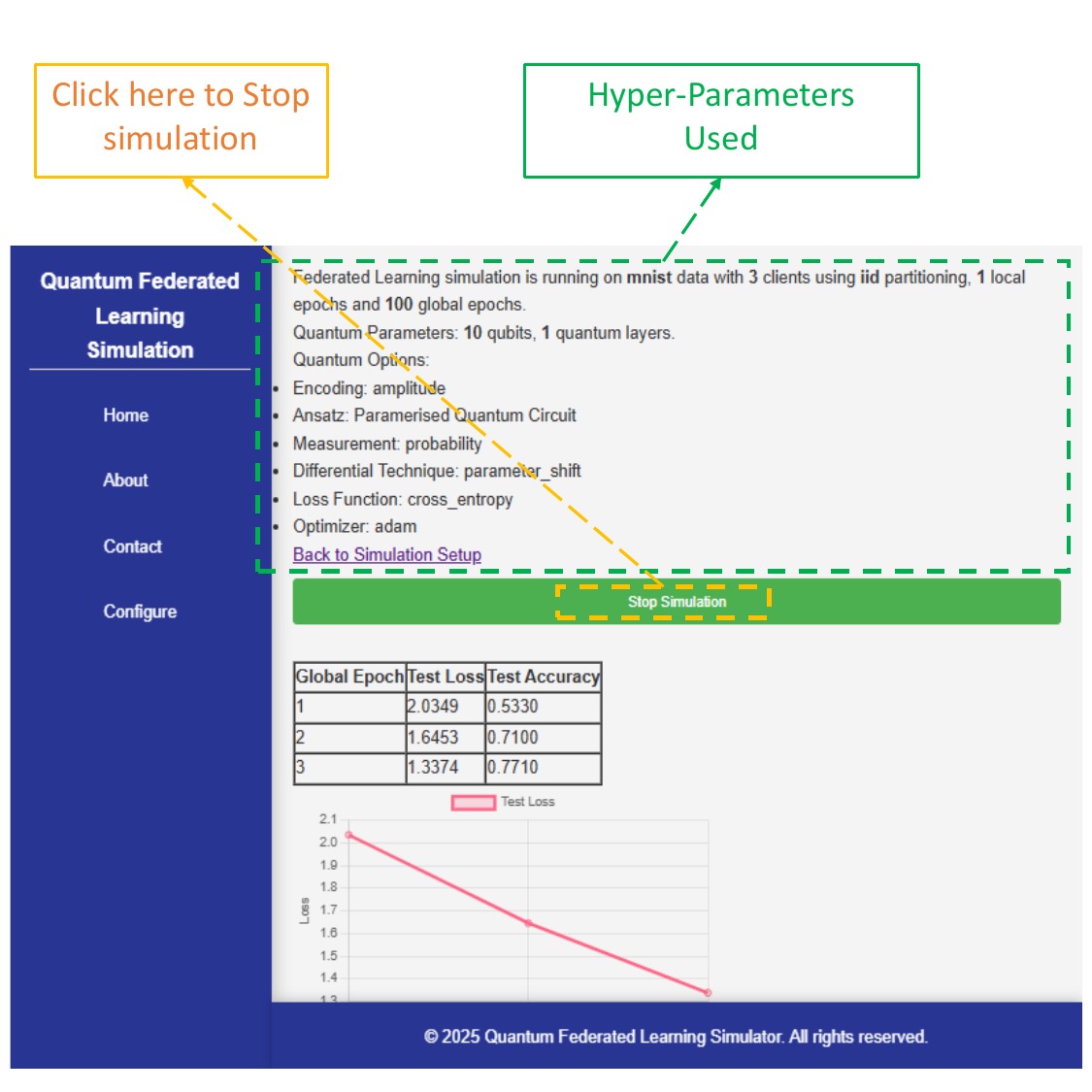}
        \caption{Simulation view (hyper-parameter).}
        \label{fig:img3}
    \end{subfigure}
    \hfill
    \begin{subfigure}[t]{0.33\linewidth}
        \centering
        \includegraphics[width=\linewidth]{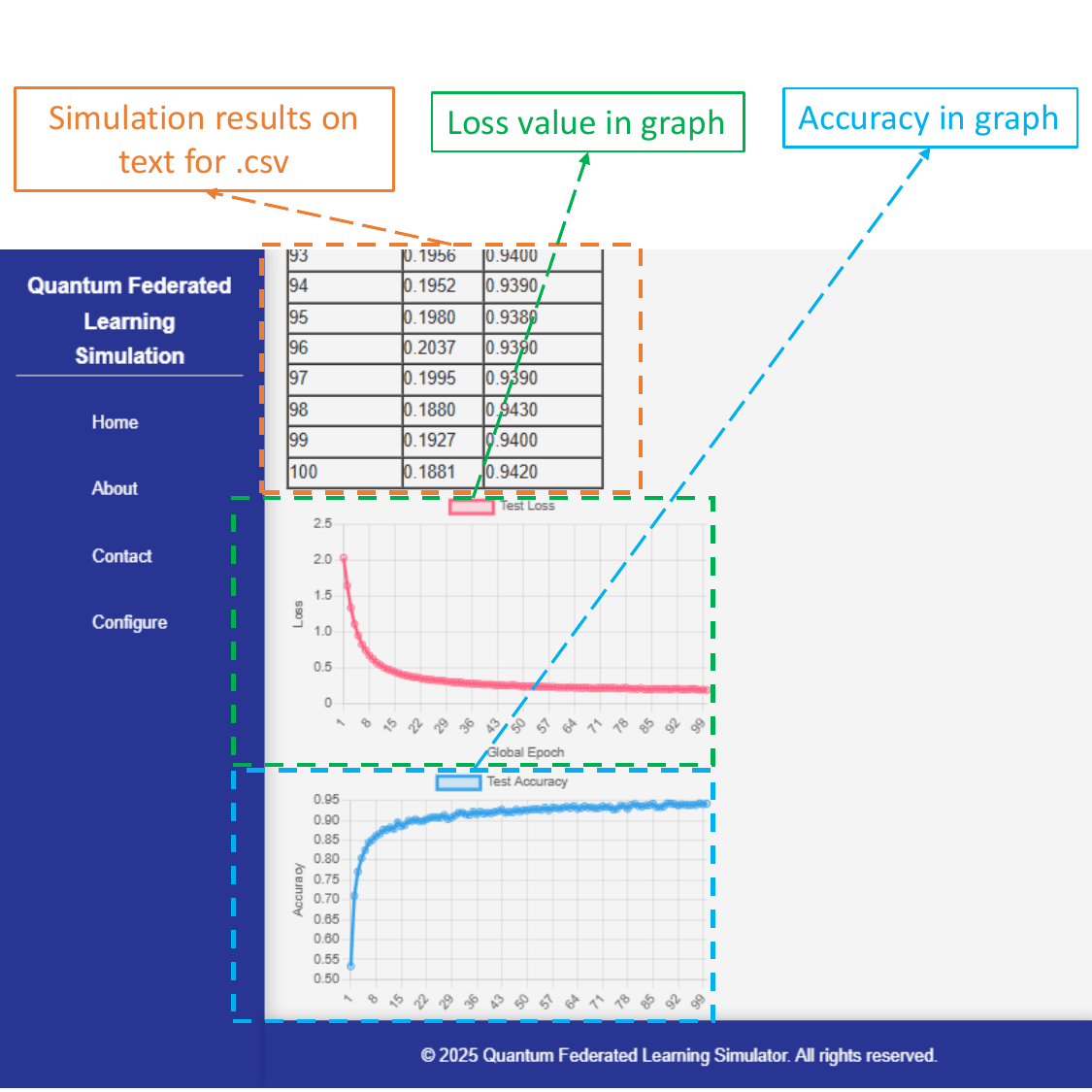}
        \caption{Simulation view (graphs and results).}
        \label{fig:img4}
    \end{subfigure}
    \hfill
    \begin{subfigure}[t]{0.33\linewidth}
        \centering
        \includegraphics[width=\linewidth]{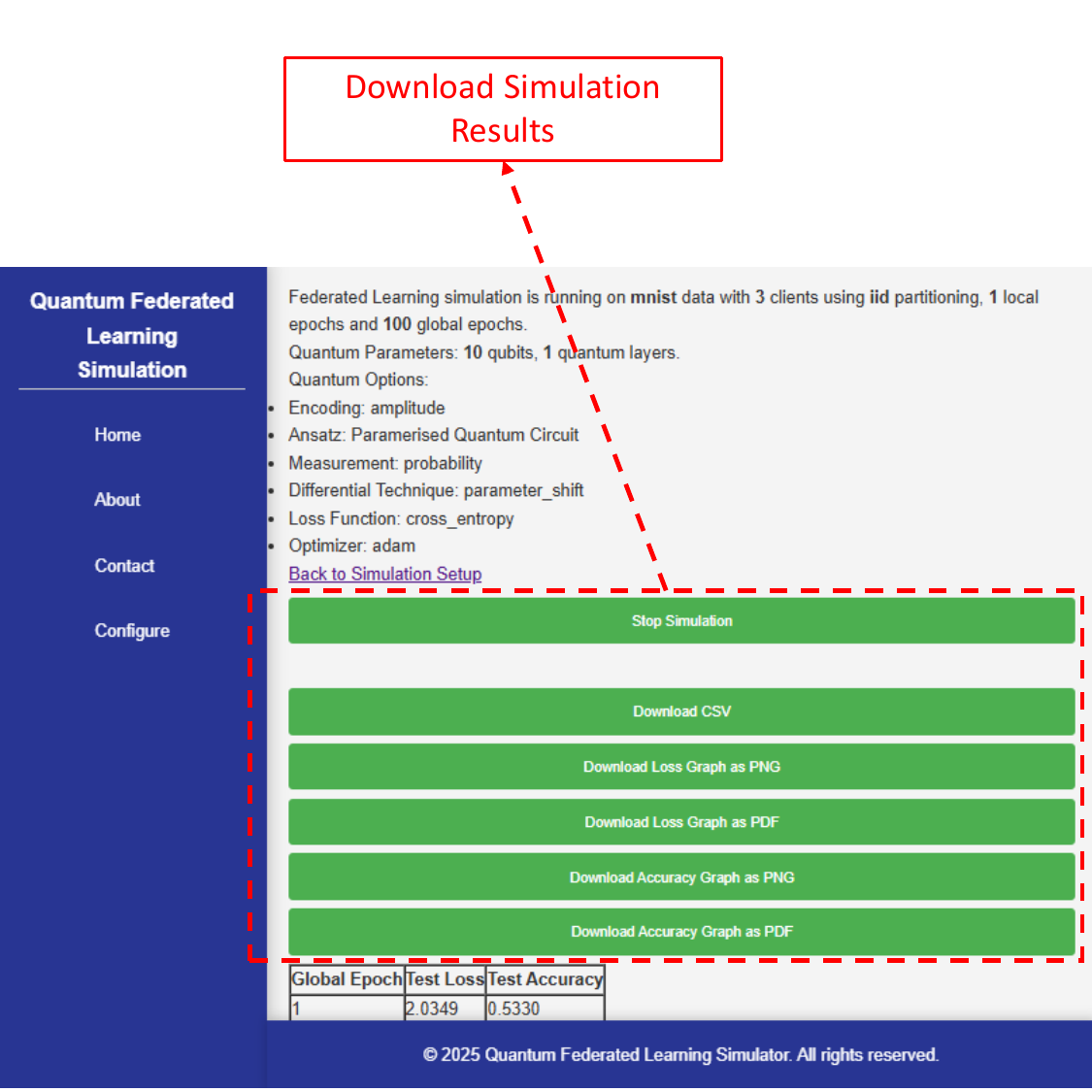}
        \caption{Simulation view (saving results).}
        \label{fig:img5}
    \end{subfigure}

    \caption{This is a simulator visualization that shows the overall structure of our simulator and the overview of the SimQFL simulator's user interface. Key navigation pathways between the various system components are depicted in the screenshots: (a) default home screen; (b) simulation with standard datasets; (c) simulation with user datasets; (d) simulation with uploaded datasets; (e) hyper-parameter settings; (f) graphical outputs; and (g) saving results.}
    \label{fig:five-grid}
\end{figure*}

To run the simulator, a main script is written in Python called \textit{app.py} is launched that initiates the simulation. Upon running the file, a web-based interface opens, and in the \textit{`Configure and Start Simulation'} menu, a user can provide the configuration values. The SimQFL simulator's visual interface and structural architecture are shown in Figure~\ref{fig:five-grid}, emphasizing its modular and interactive design.

\subsection{Part 1: Home Screen} 
Initially, we start the simulator using the $app.py$ file with a Python compiler. The journey starts on the home screen, Fig.~\ref{fig:img1}, where users start a new simulation session and are given an overview of the simulator's surroundings. The interactive and visually appealing interface allows users to navigate the entire SimQFL pipeline easily. All of the essential pages are accessible through this page.

\subsection{Part 2: Configuration Screen} 
If we select the \textit{'Configure'} button, we go to the configure page as shown in Fig.~\ref{fig:img1.5}. On that page, there are 2 options. In the first option, the user will be able to run their own standard dataset, including MNIST, FashionMNIST, and CIFAR-100, and run the simulation as shown in Fig.~\ref{fig:img2}. In the upload your data section, user can upload their data and simulate as demonstrated in Fig.~\ref{fig:img2.5}.

\subsection{Part 3: Selection of Hyperparameters}
To evaluate SimQFL's capabilities, we have created an interactive simulation environment that allows users to easily prepare and run quantum or conventional federated learning experiments.  The simulation form accepts a variety of parameters that influence both the federated learning setup and the quantum circuit architecture. Upon clicking on the \textit{Configure and Start Simulation} button, users are then taken to the simulator configuration interface Fig.~\ref{fig:img2}, \ref{fig:img2.5}, where essential training parameters are specified and chosen.

\textbf{FL Setup:} Users can select between two learning frameworks: Quantum Federated Learning (QFL) and Classical Federated Learning (FL). Although QFL is the default and most developed option in the current version, classical FL is supported and will be improved with new classical model choices. The simulator presently supports four commonly utilized benchmark datasets: MNIST \cite{deng2012mnist}, FashionMNIST \cite{xiao2017fashion}, and CIFAR-100 \cite{krizhevsky2009learning}. These datasets offer varying levels of complexity and input dimensions, allowing users to evaluate their models under a variety of scenarios. The federated learning environment is mostly customizable. Users may select: \begin{itemize}\item Client count ($n$) - up to 100 clients \item Number of local training epochs per client \item Number of global communication rounds \item Loss Function - Supports \textit{Cross Entropy Loss}, suitable for classification tasks. \item Optimizer - Employs the \textit{Adam} optimizer for both classical and quantum parameter updates. \end{itemize}
Several additional features, including support for multiple loss functions, alternative optimization algorithms, and customizable learning rates, are currently under development. These enhancements will be available in future versions for user configurations. 

\textbf{Quantum Configuration Options:} Users have access to a specific set of quantum setup settings while using the QFL.  This includes:  \begin{itemize} \item {The number of qubits ($Q_n$):}  Determines the input space dimension and circuit capacity.  \item {The number of quantum layers ($L$) is:}  Determines the depth of the parameterized quantum circuit. \end{itemize} 
The quantum encoding scheme currently only supports \textit{Amplitude Encoding}, however, other encoding techniques like angle and basis encoding will be added later. Similarly, the Ansatz Type currently supports the standard PQC entanglement gate, which is optimized for simplicity and interpretability.  The measuring method supports \textit{Probability Measurement}, with plans to incorporate expectation value- and sample-based readouts. The \textit{Parameter Shift} rule is commonly used for gradient estimation in quantum circuits with continuous parameters for differentiation techniques. 

\subsection{Part 4: Simulation View}
After clicking the \textit{Start Simulation} button, we start the simulation and real-time visualization of the result from the simulation.  When the simulation starts, users are sent to the hyperparameter visualization view Fig.~\ref{fig:img3}, where they may quickly see and confirm the simulation's settings.  The simulator continuously refreshes its visualization dashboard Fig.~\ref{fig:img4}, throughout training, providing real-time feedback in the form of convergence measures, accuracy trends, and loss curves.  This helps with the early identification of underfitting or training instability in addition to tracking model performance. Lastly, users may export trained models, graphs, and logs for external analysis or experiment reproduction in the future using the results storing interface Fig.~\ref{fig:img5}.  By doing this, simulations are guaranteed to be reusable and traceable.  SimQFL's user interface is designed with accessibility, transparency, and control in mind, making it easy for users to set up, run, monitor, and save their QFL experiments.

\textbf{Real-Time Feedback and Visualization:}
When the simulation starts, the system outputs real-time, epoch-wise feedback, displaying the testing loss and accuracy in both as a table and graphical feedback to show global model convergence. The plots are dynamically updated in every global round, enabling the users to monitor training progress, debug quantum circuit behaviors, and fine-tune hyperparameters for better performance.


\textbf{Simulation Setup:} On the \textit{backend of the simulator}, we run a Python-based application. A realistic quantum federated system is simulated by SimQFL using a number of fundamental components as follows
\begin{itemize}
    \item \textit{QuantumClient:} A decentralized node is represented by the central simulation entity.  Every QuantumClient interacts via a CommunicationLayer, encodes data using a QuantumEncoder, employs a QuantumModel, and includes a NoiseModel.
    \item \textit{QuantumModel:} a PQC that represents the client's trainable quantum neural network.
    \item \textit{QuantumEncoder:} in charge of converting conventional data into quantum states that are used in quantum computing.
    \item \textit{CommunicationLayer:} defines the protocol used for communication between the central server and QuantumClient.
    \item \textit{ClassicalServer:} arranges for client training.  After every training cycle, it updates the global model and uses an Aggregator component to aggregate model parameters.
    \item \textit{Aggregator:} gathers all client model updates, applies custom aggregation or federated averaging, and then outputs the global model.
    \item \textit{NoiseModel:} replicates actual hardware behavior by simulating the effects of noise in quantum computing.
\end{itemize}

\textbf{Simulation Lifecycle:}
\begin{figure}
    \centering
    \includegraphics[width=0.99\linewidth]{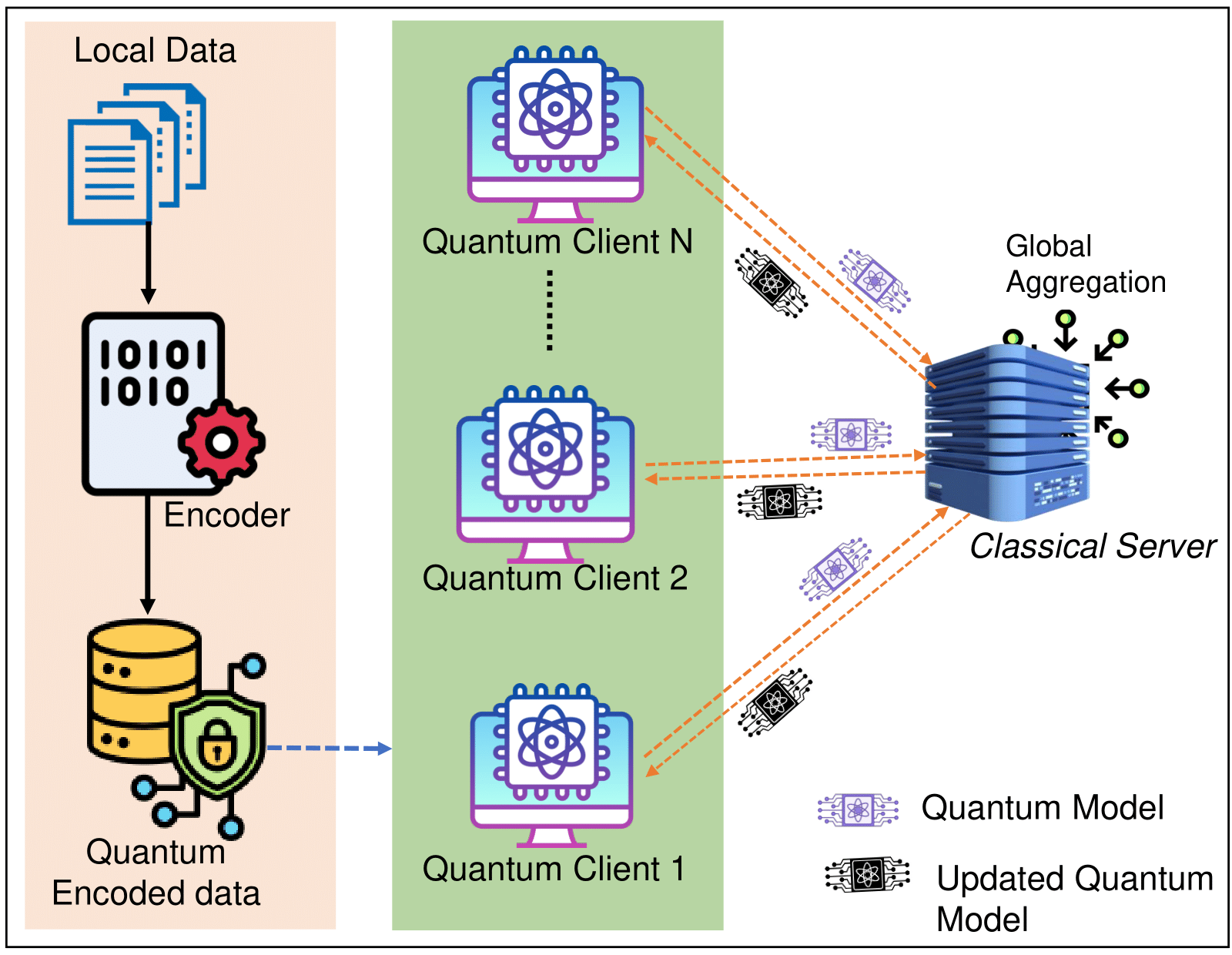}
    \caption{System diagram of quantum client's communication on the proposed QFL Simulator. Components include local data, an encoder, quantum-encoded data, a quantum model for every Quantum Client, and global aggregation on a classical server. }
    \label{fig: comm_overview}
\end{figure}

A system overview of the quantum client's communication for our proposed SimQFL approach is shown in Fig.~\ref{fig: comm_overview}. SimQFL uses a round-based simulation approach that is modified to accommodate quantum models and mimics FL workflows:
\begin{itemize}
    \item \textit{Setup phase:} The simulator generates multiple QuantumClient and initializes components such as the number of local epochs, global rounds, quantum circuit depth, and qubit count, the quantum encoding method, anstaz, measurement observable, differential technique, loss function, and learning optimizer. 
    \item \textit{Local training:} Every client updates its local setting and uses the designated data partition to carry out local quantum training using QNN.
    \item \textit{Communication phase:} Using the CommunicationLayer, clients provide the ClassicalServer with their learned parameters.
    \item \textit{Aggregation:} To create a new and updated global model, the server combines client changes using an appropriate aggregator and aggregation technique.
    \item \textit{Broadcast:} For the subsequent round, the server provides the clients with the updated model.
    \item \textit{Visualization:} After every round, the simulator displays updated accuracy/loss charts, graphs, and convergence behavior.

\end{itemize}

\section{Use Case}
In order to assess our simulator, we developed a QFL algorithm that uses a simulated quantum backend to show the distributed training of QNN models over multiple clients.  Each client uses PQC, made up of variational quantum layers, to train its local QNN model on a subset of the dataset.  A quantum-aware version of the Federated Averaging (FedAvg) technique, which enables hybrid classical-quantum weight structures, is used to aggregate model parameters following local training.

We configure SimQFL to simulate a scenario with $n$ quantum devices (clients) selected by the user and a centralized server. Each device uses QML to train its local model.  Over multiple global rounds, a QNN model and stochastic gradient descent (SGD) are used for training.  Each quantum device performs numerous local training epochs during each global cycle.  The dataset utilized during a given round is considered local to that device.   This data is converted to quantum format for processing.   The datasets from all devices are combined for each cycle, and the data distributions are non-uniform, with each device processing a different amount of data.   During each global epoch, each device chooses a random data sample from its dataset for training.  The Adam optimizer adjusts the learning rate as needed to maintain control over the training process. Following training, the server gathers and aggregates updates from all devices.  The aggregated global model is then sent to all devices for the following training rounds. After each cycle, test data is used to assess performance. SimQFL includes epoch-wise representations of local client loss, global accuracy, and convergence trends, all of which are updated in real time.  These representations allow the user to watch each client's learning behavior, evaluate the contribution of quantum layers to performance, and immediately discover abnormalities or bottlenecks in the training loop.


\begin{algorithm}
\footnotesize
    \caption{Quantum Federated Learning (QFL) Algorithm}
    \label{algo:qfl}
    \begin{algorithmic}[1]
        \State \textbf{Input:} Number of global rounds \( R \), local epochs \( K \), clients \( \mathcal{N} = \{1, \dots, N\} \), learning rate \( \eta \), number of shots \( M \)
        \State \textbf{Initialization:} Initialize global model parameters \( \boldsymbol{\omega}_{\text{global}}^{(0)} \)
        \For{each global round \( r = 1, \dots, R \)}
            \State Server broadcasts \( \boldsymbol{\omega}_{\text{global}}^{(r)} \) to all clients
            \For{each client \( n \in \mathcal{N} \) in parallel}
                \State Set local parameters: \( \omega_{n,r}^{0} = \boldsymbol{\omega}_{\text{global}}^{(r)} \)
                \For{each local epoch \( k = 0, \dots, K{-}1 \)}
                    \State Sample mini-batch \( B_n^k \subseteq D_n \)
                    \For{each data point \( (w, y) \in B_n^k \)}
                        \State Encode input using amplitude encoding \eqref{eq:encoding_amplitude}
                        \State Apply PQC to encoded state \eqref{eq:pqc_output_state}
                        \State Compute prediction via measurement expectation \eqref{eq:measurement_expectation}
                        \State Evaluate sample loss \( \ell(y, \hat{f}_{n,r}^{k}) \)
                    \EndFor
                    \State Compute gradients using parameter shift rule \eqref{eq:parameter_shift}
                    \State Update model parameters using gradient descent \eqref{eq:parameter_update}
                \EndFor
                \State Send updated parameters \( \omega_{n}^{(r)} \) to server
            \EndFor
            \State Server aggregates client updates via weighted averaging \eqref{eq:aggregation}
            \State Server broadcasts \( \boldsymbol{\omega}_{\text{global}}^{(r+1)} \) to all clients
        \EndFor
        \State \textbf{Output:} Final global model \( \boldsymbol{\omega}^* = \boldsymbol{\omega}_{\text{global}}^{(R)} \)
    \end{algorithmic}
\end{algorithm}

Algorithm~\ref{algo:qfl} summarizes the QFL framework. It begins with global model initialization (line~1), followed by $R$ global rounds (lines~2--17). In each round, the server broadcasts the global model (line~4), and clients perform local training on mini-batches (lines~5--13). This includes amplitude encoding (line~9), PQC application and measurement (line~10), and parameter updates using the rule in Eq.~\eqref{eq:parameter_update} (line~12). Clients send updated models to the server (line~13), which aggregates them using Eq.~\eqref{eq:aggregation} (line~15). The process repeats until round $R$.

\begin{figure}
    \centering
    \footnotesize
    \begin{subfigure}[t]{0.52\linewidth} 
        \centering
        \includegraphics[width=\linewidth]{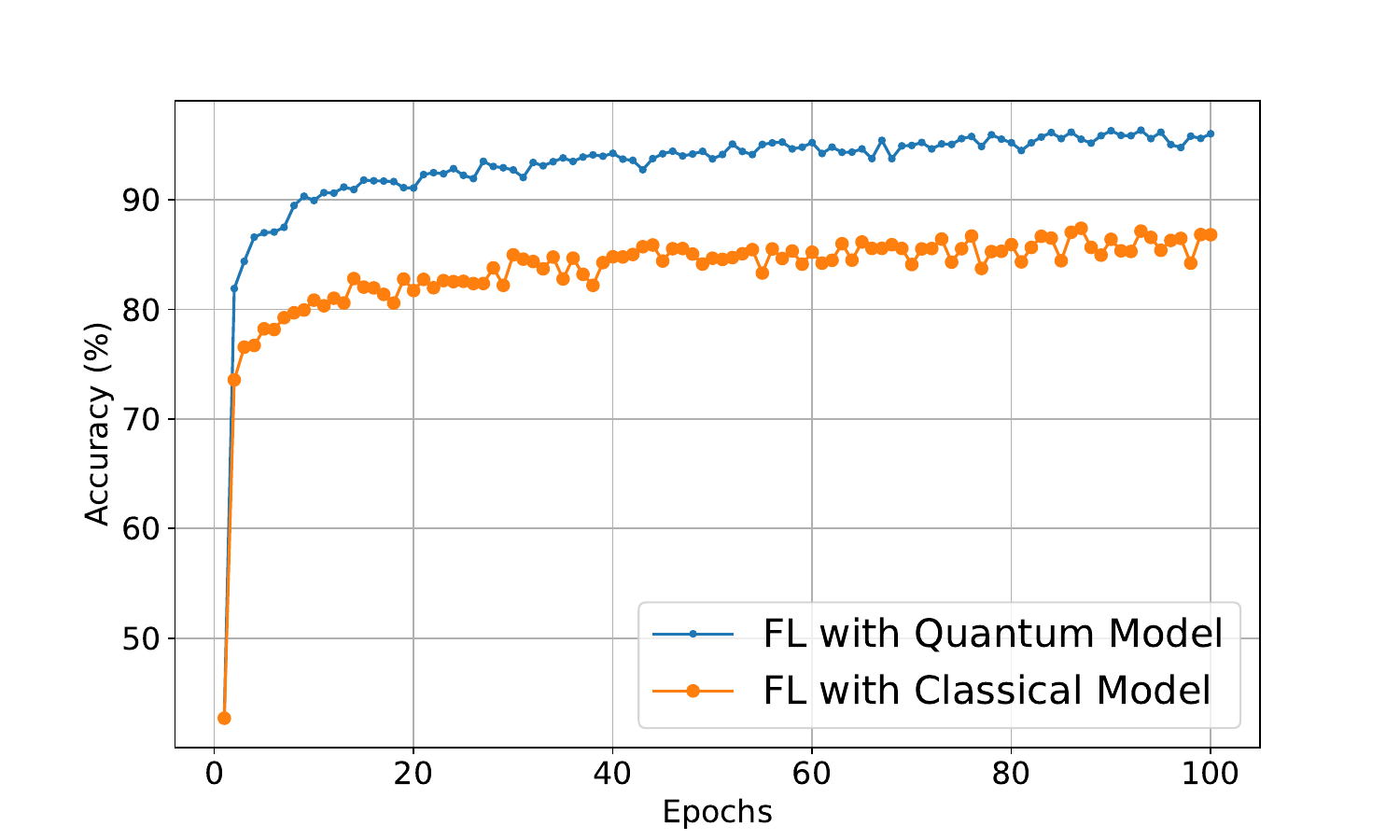}
        \caption{MNIST accuracy (\%)}
        \label{fig2a}
    \end{subfigure}
    \hspace{-2cm}
    \hfill 
    \begin{subfigure}[t]{0.52\linewidth} 
        \centering
        \includegraphics[width=\linewidth]{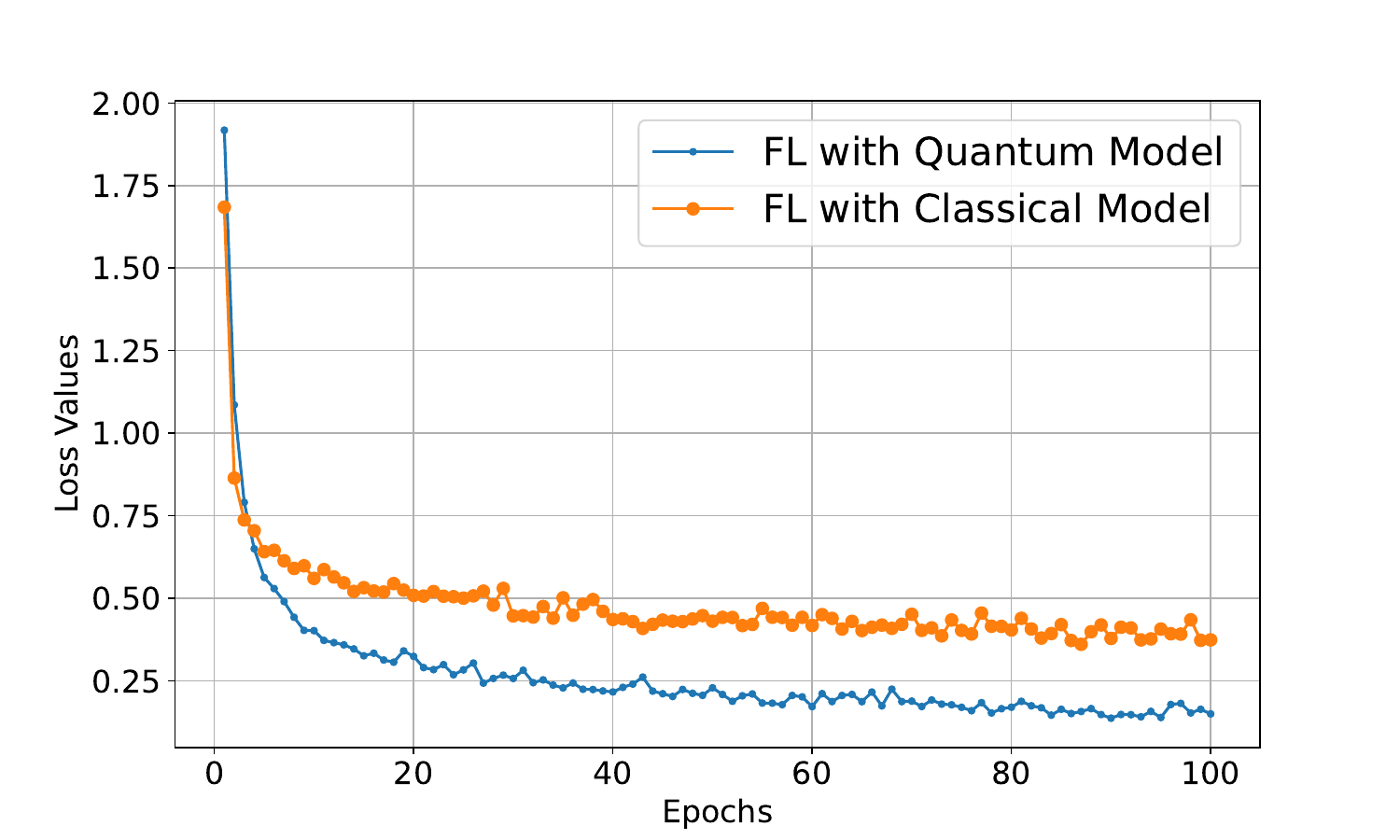}
        \caption{MNIST loss}
        \label{fig2b}
    \end{subfigure}
    \begin{subfigure}[t]{0.52\linewidth} 
        \centering
        \includegraphics[width=\linewidth]{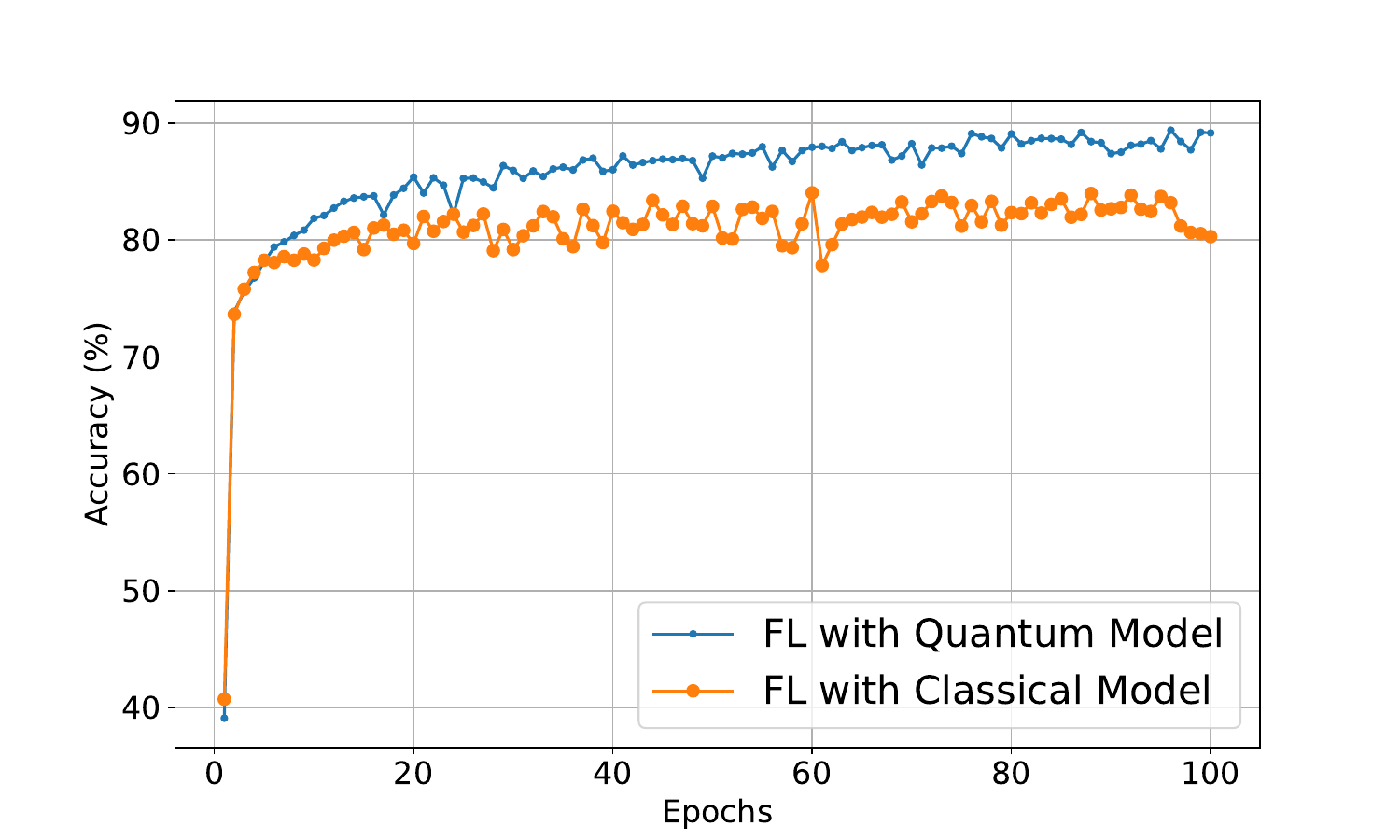}
        \caption{Fashion-MNIST accuracy (\%)}
        \label{fig2c}
    \end{subfigure}
    \hspace{-2cm}
    \hfill 
    \begin{subfigure}[t]{0.52\linewidth} 
        \centering
        \includegraphics[width=\linewidth]{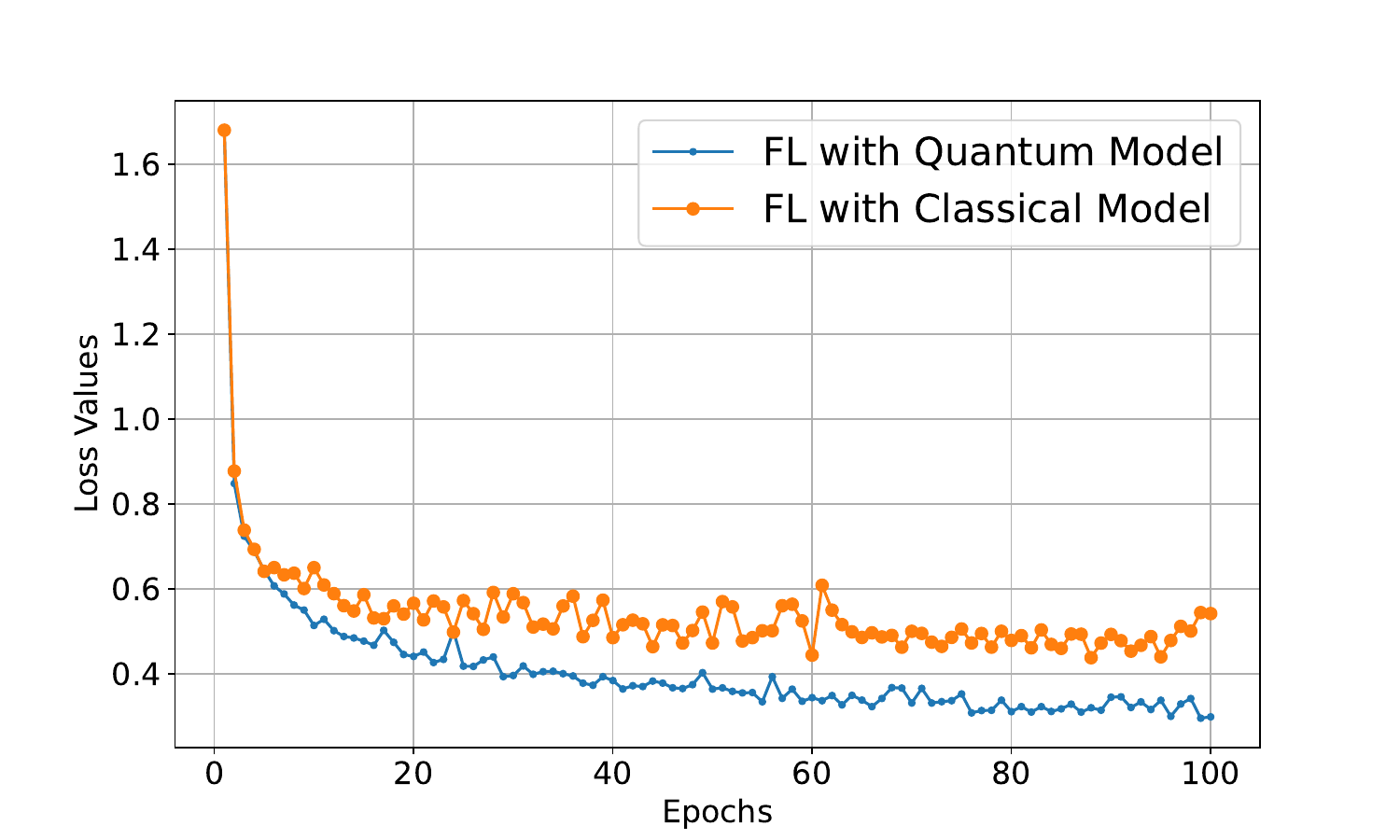}
        \caption{Fasion-MNIST loss}
        \label{fig2d}
    \end{subfigure}
    \begin{subfigure}[t]{0.52\linewidth} 
        \centering
        \includegraphics[width=\linewidth]{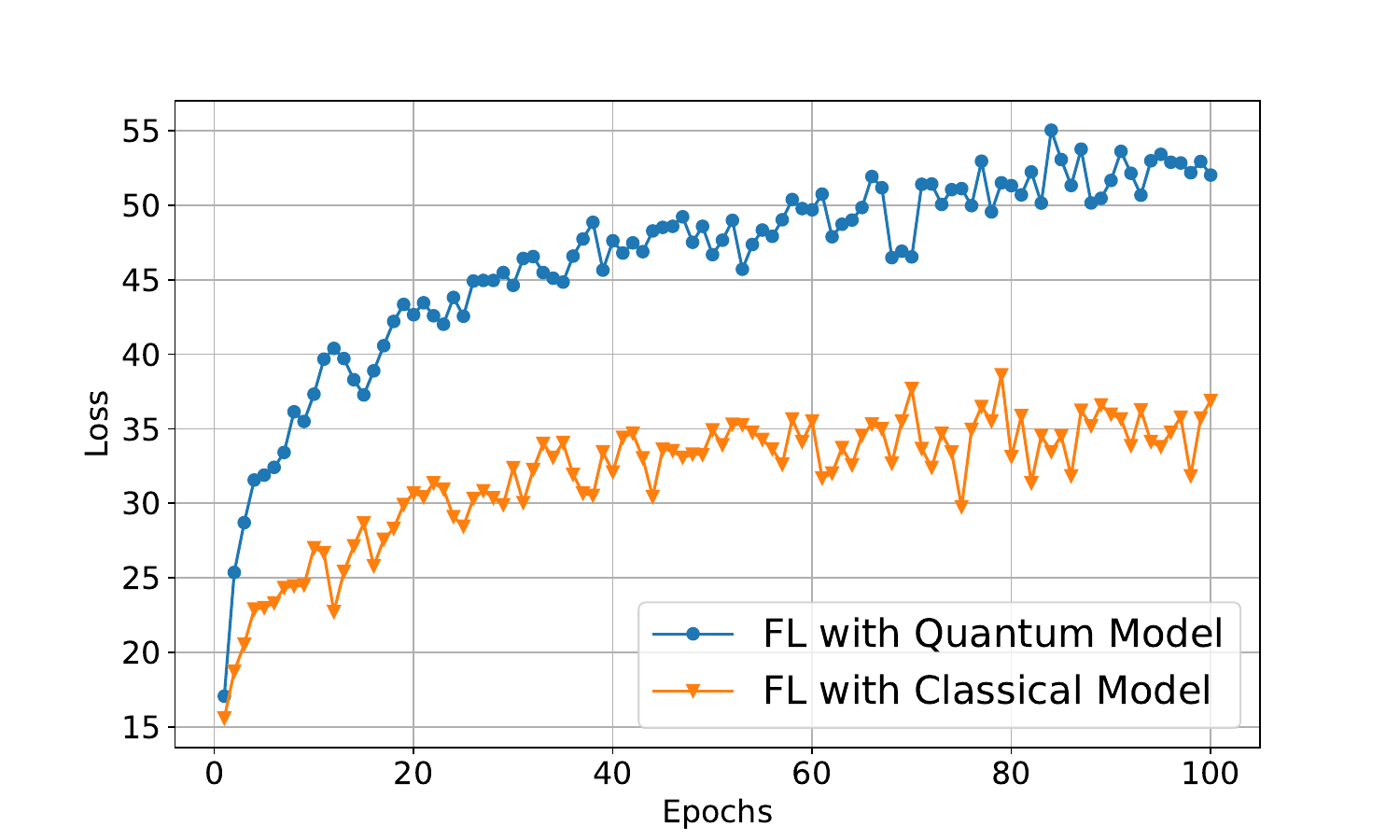}
        \caption{CIFAR-100 accuracy (\%)}
        \label{fig2e}
    \end{subfigure}
    \hspace{-2cm}
    \hfill 
    \begin{subfigure}[t]{0.52\linewidth} 
        \centering
        \includegraphics[width=\linewidth]{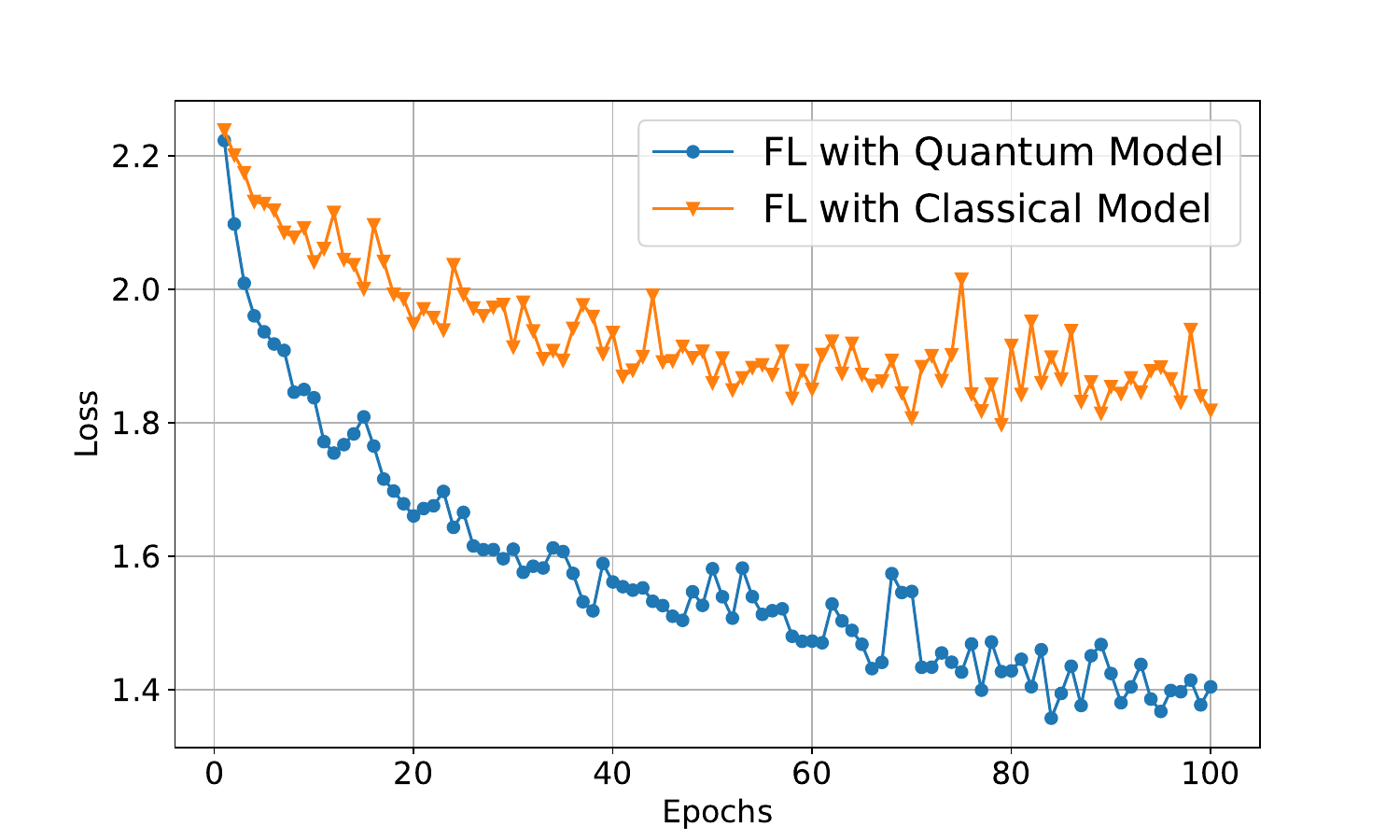}
        \caption{CIFAR-100 loss}
        \label{fig2f}
    \end{subfigure}
    \caption{Performance comparison between FL with quantum model (QFL), and FL with classical model (classical FL) approaches. We use MNIST, Fashion-MNIST, and CIFAR-100 datasets for comparison.}
    \label{fig1}
\end{figure}

This use case shows how SimQFL makes experimentation in quantum federated learning simpler by providing high configurability, intuitive visualization, and seamless simulation support for hybrid quantum-classical procedures.  This allows researchers to quickly build QFL algorithms, test hyperparameter settings, and acquire a better understanding of the dynamics of distributed quantum learning.

\section{Experiments}

\subsection{Effects on Performance Using QFL Over FL:}
To evaluate the effects of using the quantum mode, we plot the performance comparison of both loss and accuracy values in MNIST, FashionMNIST, and CIFAR-100 data as shown in Fig.~\ref{fig1}. The graphs show that FL with the Quantum model has outperformed FL with the Classical model in terms of both prediction and convergence. QFL has outperformed the classical FL in the MNIST data around 10.5\% in Fig.~\ref{fig2a}, FashionMNIST data around 7.3\% in Fig.~\ref{fig2b}, and CIFAR-100 data around 23.1\% in Fig.~\ref{fig2c} in terms of overall accuracy. Also, we can see that in QFL, the accuracy has been achieved significantly faster than traditional FL, showing a better convergence speed.

\subsection{Accuracy vs Qubits Tradeoff:}
Using different qubit values in the simulator gives us an overall idea of the effects of qubits on the overall accuracy. We selected the qubit values of 1, 2, 5, and 10 for comparison.  The experimental findings are summarized in Table \ref{tab:qubits}.  All settings, including training epochs and hyperparameters, remained consistent across experiments, except the number of qubits, denoted as $Q_n$.  We evaluated $Q_n$ values of 2, 3, 5, and 10 to assess the effect of quantum capacity.  The QFL model performed best with $Q_n = 10$, indicating an advantageous relationship between the qubit count and overall performance. Due to its complex figures and high-dimensional image data (32 x 32 x 3 pixels), CIFAR-10 cannot operate on a single qubit. Also, the results indicate that we would get better performance with more qubits. However, 10 qubits require a substantial amount of processing power and resources, so we have chosen not to add additional qubits. As a result, all future simulations used the default setup of 10 qubits. 

\begin{table*}
    \centering
    \footnotesize
    \caption{Difference in performance in quantum learning with the different number of qubits across various datasets.}
    \label{tab:qubits}
    \begin{tabular}{|c|cc|cc|cc|}
    \toprule
    \multicolumn{1}{|c|}{\multirow{2}{*}{N (Num Qubit)}} & \multicolumn{2}{c|}{\textbf{MNIST}} & \multicolumn{2}{|c|}{\textbf{FashionMNIST}} & \multicolumn{2}{c|}{\textbf{CIFAR-100}} \\
    \cmidrule{2-7}
    & \textbf{Loss Value} & \textbf{Acc.} & \textbf{Loss Value} & \textbf{Acc.} & \textbf{Loss Value} & \textbf{Acc.} \\
    \midrule
    1 & 2.311 & 9.88\% & 2.32 & 9.83\% & X & X \\
    2 & 2.108 & 20.68\% & 2.212 & 15.93\% & 4.026 & 11.54\% \\
    5 & 1.872 & 43.33\% & 1.667 & 33.87\% & 3.011 & 29.78\% \\
    10 & \textbf{0.109} & \textbf{97.13\%} & \textbf{0.283} & \textbf{90.33\%} & \textbf{1.320} & \textbf{55.04\%} \\
    \bottomrule
    \end{tabular}
\end{table*}

\subsection{Accuracy vs Quantum Layers Tradeoff}
We examine the performance of various quantum layers by changing the number of layers $L$ and assessing the resulting QFL performance on the MNIST, FashionMNIST, and CIFAR-100 datasets.  Table~\ref{tab:layer} demonstrates that using fewer quantum layers (especially $L$=1 and $L$=2) yields the optimal balance of accuracy and loss across all datasets.  Increasing the number of layers beyond this threshold results in a constant drop in performance, most likely due to overparameterization and optimization problems like vanished gradients or barren plateaus.  Notably, MNIST reaches a peak accuracy of 97.13\% at $L$=2, but FashionMNIST and CIFAR-100 reach peak accuracy at $L$ = 1 of 91.11\% and 55.63\%, respectively, emphasizing the necessity of selecting an ideal quantum depth based on dataset complexity and model capacity.

\begin{table*}
    \centering
    \caption{Difference in performance in quantum learning with the different number of layers across various datasets.}
    \label{tab:layer}
    \begin{tabular}{|c|cc|cc|cc|}
    \toprule
    \multicolumn{1}{|c|}{\multirow{2}{*}{$L$ (num layers)}} & \multicolumn{2}{c|}{\textbf{MNIST}} & \multicolumn{2}{|c|}{\textbf{FashionMNIST}} & \multicolumn{2}{c|}{\textbf{CIFAR-100}} \\
    \cmidrule{2-7}
    & \textbf{Loss Value} & \textbf{Acc.} & \textbf{Loss Value} & \textbf{Acc.} & \textbf{Loss Value} & \textbf{Acc.} \\
    \midrule
    1 & 0.1732 & 96.20\% & \textbf{0.2724} & \textbf{91.11\%} & \textbf{1.3388} & \textbf{55.63\%} \\
    2 & \textbf{0.1612} & \textbf{97.13\%} & 0.2841 & 90.48\% & 1.3716 & 53.18\% \\
    3 & 0.2009 & 95.01\% & 0.2991 & 89.18\% & 1.3660 & 52.98\% \\
    4 & 0.1950 & 95.32\% & 0.3397 & 88.17\% & 1.4045 & 52.02\% \\
    5 & 0.3319 & 91.05\% & 0.3864 & 86.08\% & 1.4935 & 48.34\% \\
    10 & 0.4211 & 86.38\% & 0.5421 & 80.30\% & 1.8187 & 36.89\% \\
    \bottomrule
    \end{tabular}  
\end{table*}
\begin{table*}
    \centering
    \caption{Performance comparison in quantum learning with different numbers of clients across various datasets (highlighting optimal performance for 10 clients).}
    \label{tab:clients}
    \begin{tabular}{|c|cc|cc|cc|}
    \toprule
    \multicolumn{1}{|c|}{\multirow{2}{*}{$C$ (Num Clients)}} & \multicolumn{2}{c|}{\textbf{MNIST}} & \multicolumn{2}{|c|}{\textbf{FashionMNIST}} & \multicolumn{2}{c|}{\textbf{CIFAR-100}} \\
    \cmidrule{2-7}
    & \textbf{Loss Value} & \textbf{Acc.} & \textbf{Loss Value} & \textbf{Acc.} & \textbf{Loss Value} & \textbf{Acc.} \\
    \midrule
    2 & 0.3319 & 91.05\% & 0.3864 & 86.08\% & 1.4935 & 48.34\% \\
    3 & 0.2009 & 95.01\% & 0.2991 & 89.18\% & 1.3660 & 52.98\% \\
    5 & 0.1612 & 97.13\% & 0.2841 & 90.48\% & 1.3716 & 53.18\% \\
    10 & \textbf{0.1612} & \textbf{97.13\%} & \textbf{0.2724} & \textbf{91.11\%} & \textbf{1.3388} & \textbf{55.63\%} \\
    \bottomrule
    \end{tabular}  
\end{table*}

\subsection{Accuracy vs Number of Clients Tradeoff}
To examine the effect of client involvement on QFL performance, we adjust the number of clients $C$ and test the model using three benchmark datasets: MNIST, FashionMNIST, and CIFAR-100. Table~\ref{tab:clients} shows that increasing the number of clients increases both accuracy and loss. With a comparable loss decrease from 0.3319 to 0.1612, MNIST's accuracy increases dramatically from 91.05\% at \( C=2 \) to 95.01\% at \( C=3 \), peaking at 97.13\% at \( C=5 \) and remaining constant at \( C=10 \), suggesting particularly in the early phases, the model profits from distributed updates and client variety. FashionMNIST's performance also steadily improves as the number of clients increases: accuracy increases from 86.08\% at \( C=2 \) to 89.18\% at \( C=3 \), 90.48\% at \( C=5 \), and reaches its highest point at 91.11\% with \( C=10 \).  As customer involvement rises, the loss value concurrently decreases from 0.3864 to 0.2724, indicating more consistent convergence. The pattern is less apparent but still continuous with the more complicated CIFAR-100 dataset.  With the loss decreasing from 1.4935 to 1.3388, accuracy rises from 48.34\% at \( C=2 \) to 52.98\% at \( C=3 \), 53.18\% at \( C=5 \), and achieves its maximum value of 55.63\% at \( C=10 \).  These enhancements highlight the advantages of FL, especially in situations involving high-dimensional data, where the global quantum model benefits from cooperative updates from many clients. Overall, the findings indicate that more clients improve model performance, with C=10 consistently showing the best accuracy and lowest loss across all datasets.

\section{Availability}
To encourage repeatability, openness, and practicality in Quantum Federated Learning (QFL) research, we have made our developed simulator accessible to everyone via GitHub at \url{https://github.com/Ratun11/SimQFL}.  The simulator is delivered as a compressed archive entitled \texttt{simulator.zip}. This archive comprises a self-contained, pre-compiled Linux executable (\texttt{app}) produced using PyInstaller.  This packing strategy provides smooth deployment on any x86-64 Linux machine without needing the installation of Python or other dependencies.  The simulator has been tested on Ubuntu 22.04 and is likely to work consistently with other current Linux versions.

After extracting the package, users may start the simulator by running the binary file from the terminal.  The simulator launches a Flask-based web server that may be visited via \url{http://127.0.0.1:5000} in any contemporary web browser.  Users may utilize the interactive interface to execute decentralized training simulations by adjusting the number of clients, quantum layers, qubits, encoding methods, learning rates, and other essential factors.  The simulator supports numerous common datasets, including MNIST, FashionMNIST, CIFAR-10, and CIFAR-100, as well as the ability to upload new datasets in CSV format.

The interface also includes real-time performance visualization and the ability to export simulation results for additional examination.  This binary distribution method provides not just widespread accessibility and ease of use, but also the security of the simulator's source code.  We anticipate that our open-access simulator will be a helpful tool for academics investigating QFL architectures and protocols, laying the framework for future development of privacy-preserving, quantum-enhanced learning systems.

\section{Conclusion and Future Work}

This study introduced SimQFL, a lightweight and powerful simulator that bridges the gap between quantum machine learning and federated learning by offering a single platform for Quantum Federated Learning (QFL).  SimQFL fills three major holes in current systems: the lack of integrated quantum-federated training settings, the lack of intuitive, real-time visualizations, and the inability to train on bespoke, user-uploaded data.  SimQFL makes it easier for researchers to investigate and assess QFL approaches by providing support for quantum encoding, variational quantum layers, client-specific configurations, and real-time visibility into training dynamics. SimQFL's standalone executable deployment methodology offers maximum accessibility and system independence without affecting privacy. This deployment technique makes it ideal for instructional, experimental, and practical research applications.  Researchers may experiment with both common datasets (e.g., MNIST, FashionMNIST, and CIFAR-10) and user-defined data in a variety of federated setups, changing qubit counts and quantum circuit depths. The release of SimQFL offers a strong and adaptable platform for QFL research. The simulator's front end provides live visualizations of hyperparameters and training progress, such as loss curves, accuracy trends, and convergence. Furthermore, users may export models, metrics, and graphs for external review and reproducibility. These features make SimQFL a useful and extendable tool for evaluating QFL algorithms under realistic conditions. 

In future development, we want to considerably enhance the simulator's capability.  We will improve conventional federated learning support by introducing various model foundations and optimization methodologies.  Alternative encoding systems (like angle encoding), a variety of ansatz designs, and solely quantum optimization algorithms like QAOA are examples of quantum-side enhancements that we plan to add to this simulation.  Furthermore, we want to provide more flexible user data integration by enabling file formats other than CSV.  The simulator will also include quantum noise models and error mitigation approaches to emulate more realistic settings, improving its practicality for near-term quantum hardware. Finally, we want to incorporate noise models and error mitigation techniques to improve the simulator's applicability to near-term quantum devices. Overall, SimQFL serves an important role in advancing research in both conventional and quantum worlds by allowing for scalable, cost-effective, and repeatable experiments. 

\bibliography{main}
\bibliographystyle{IEEEtran}

\end{document}